\documentclass[aps,twocolumn,prb,showpacs,amsmath,amssymb,floatfix]{revtex4-1}

\usepackage{etoolbox} 
\usepackage{xspace}
\usepackage{graphicx}
\usepackage{dcolumn}
\usepackage{float}
\usepackage[tight]{subfigure}
\usepackage{amsmath}
\usepackage{amssymb} 
\usepackage{verbatim}
\usepackage{units}
\usepackage[usenames,dvipsnames]{color}
\usepackage{bm}
\usepackage[normalem]{ulem}
\usepackage[capitalize]{cleveref}

\newrobustcmd{\txt}[1]{\textcolor{blue}{#1}\xspace}
\newrobustcmd{\txtred}[1]{\textcolor{red}{#1}\xspace}
\newrobustcmd{\key}[1]{\textcolor{OliveGreen}{#1}\xspace}
\newrobustcmd{\landauerS}{$k_\text{B} \ln 2$~}
\newrobustcmd{\landauer}{$k_\text{B} T\ln 2$~}
\newrobustcmd{\taurel}{$\tau_\text{rel}$}
\newrobustcmd{\kb}{$k_\text{B}$~}

\renewcommand{\vec}[1]{\bm{#1} }

\newcommand{\SuppInfo}{Appendix~}

\newcommand{\Eq}[1]{equation~(\ref{#1})}

\newcommand{\Fig}[1]{Fig.~\ref{#1}}
\newcommand{\Figure}[1]{Figure~\ref{#1}}

\newcommand{\Ref}[1]{Ref.~\onlinecite{#1}}

\newcommand{\FigS}[1]{Fig.~\ref{#1}}

\begin{document}
\title{Quantum-enhanced Landauer erasure and storage}


\author{R. Gaudenzi$^{1}$}
\email{r.gaudenzi@tudelft.nl}
\author{E. Burzur\'{\i}$^{1}$}
\author{S. Maegawa$^{2}$}
\author{H. S. J. van der Zant$^{1}$}
\author{F. Luis$^{3}$}
\affiliation{$^{1}$Kavli Institute of Nanoscience, Delft University of Technology, 2600 GA, Delft, The Netherlands}
\affiliation{$^{2}$Graduate School of Human and Environmental Studies, Kyoto University, Kyoto 606-8501, Japan}
\affiliation{$^{3}$Instituto de Ciencia de Materiales de Arag\'{o}n (ICMA), C.S.I.C.-Universidad de Zaragoza, E-50009 Zaragoza, Spain}

\maketitle

\textbf{The erasure of a bit of information encoded in a physical system
is an irreversible operation bound to dissipate an amount of energy $Q=$~\landauer \cite{Landauer1961}.  
As a result, work $W \geq Q$ 
has to be applied to the physical system
to restore the erased information content \cite{Bennett1982, Bennett1988, Leff1990}. 
This limit, called Landauer limit, sets a minimal energy dissipation inherent to any classical computation.  
In the pursuit of the fastest and most efficient means of computation, the ultimate challenge is to produce
a memory device executing an operation as close to this limit in the shortest time possible. 
Here, we use a crystal of molecular nanomagnets as a spin-memory device and measure the work needed to carry out a storage operation. 
Exploiting a form of quantum annealing,
we border the Landauer limit while preserving fast operation. Owing to the tunable and fast dynamics of this process, the performance of our device in terms of energy-time cost is orders of magnitude better than existing memory devices to date. This result suggests a way to enhance classical computations by using quantum processes. }

While a computation performed with an ideal binary logic gate (e.g. NOT) 
has no lower energy dissipation limit \cite{Lambson2011, Gammaitoni2015}, 
one carried out in a memory device does.  	
The reason is that in the former the bit is merely \emph{displaced} isentropically in the space of states, whereas in the latter the minimal operation comprises an entropy non-conserving erasure-storage cycle.
In the \emph{erasure} step, the bit is allowed to \emph{explore} the two binary states and the phase space doubles with a consequent entropy increase of $\Delta S = k_\text{\text{B}} \ln 2$. 
A corresponding minimal dissipated heat \mbox{$Q = k_\text{B} T\ln 2$}, called the Landauer limit, results from this entropy change.  
In the \emph{storage} step, a work $W \geq Q$ has to be applied to \emph{reduce} the system's entropy and phase space to their initial values.     
In order to reach the $W = Q$ limit, \emph{reversible} operation is required. 
This condition is fulfilled only when using a frictionless system in a quasi-static fashion, i.e., at timescales slower than its relaxation time \taurel, so that unwanted memory and hysteresis effects are avoided.
For this reason, slower (faster) operation is generally associated with a lower (higher) dissipation. 
\begin{figure}
\centering
\includegraphics[width=.99\columnwidth]{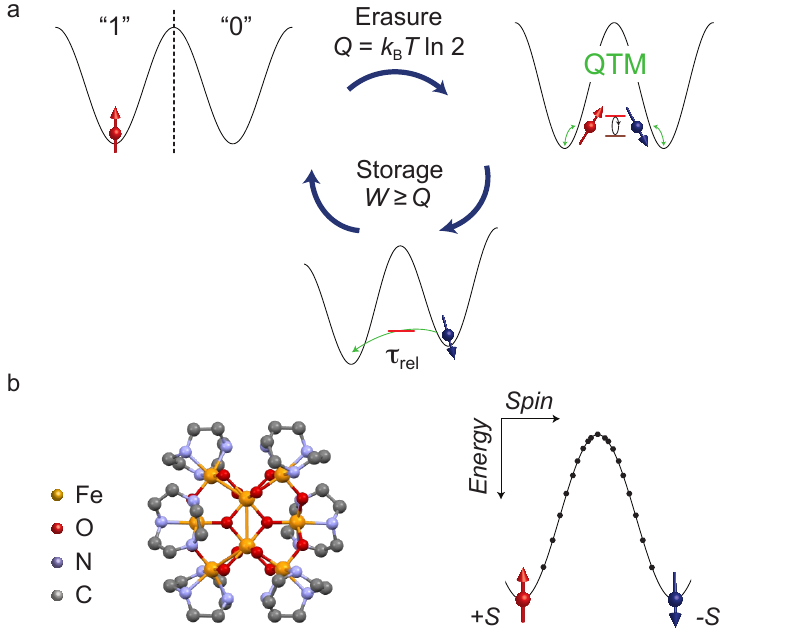}
  \caption{
  \textbf{Quantum-enhanced Landauer erasure and storage of a molecular bit.} 
     (a) Schematics of the Landauer erasure process employing a quantum nanomagnet. 
In order to erase the spin bit, the effective barrier separating the two binary states is lowered by inducing quantum tunneling of magnetization (QTM). A small bias magnetic field is then used to initialise the spin in the desired state within a time \taurel~and store the new information. The Landauer principle fixes the minimal dissipated heat $Q$ and work $W$ involved in the cycle.  
(b) Sketch of the Fe$_8$ easy-axis molecular magnet. In the absence of magnetic field, the double-well potential favors the two $S_z = \pm 10$ easy-axis spin eigenstates. 
}
\label{Fig_1}
\end{figure}
\begin{figure}
\centering
\includegraphics[width=.99\columnwidth]{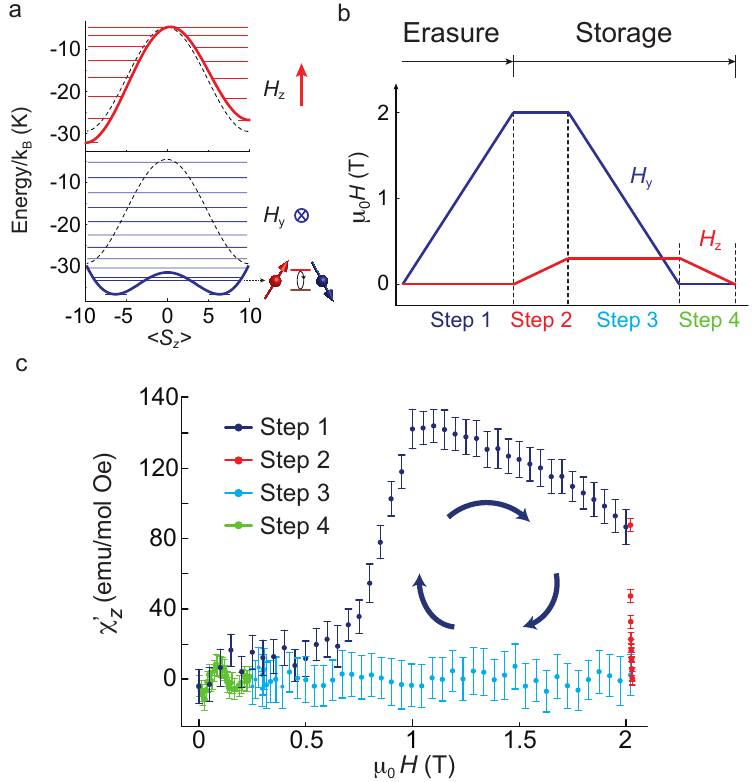}
  \caption{
  \textbf{Susceptibility of the quantum MM along the erasure-storage cycle.}
(a) Magnetic energy of a Fe8 MM subject to a $\mu_{0}H_{z} = 0.2$ T longitudinal magnetic field (top) and a $\mu_{0}H_{y} = 2 $ T transverse magnetic field (bottom). Thick solid lines show the classical potential landscape while thin horizontal lines give the quantum energy levels determined by numerical diagonalisation of Eq. (1). The dotted lines show the $H=0$ potential. While $H_{z}$ introduces an energy bias between up and down spin states, thus increasing the magnetic polarization along the easy axis, $H_{y}$ keeps the symmetry of the potential intact but promotes QTM between these states.
(b) The 4-step sequence of magnetic fields $H_y$ (blue) and $H_z$ (red) constituting the erasure-storage algorithm. The first step ($H_y: 0 \rightarrow 2$ T) consists of the Landauer erasure and the remaining 3 steps correspond to the storage protocol.
(c) Real component of the longitudinal magnetic ac-susceptibility, $\chi'_{z}$, as a function of the vector magnetic field's modulus, measured at 1K and a frequency of 333 Hz. The sequence of steps corresponds to the one in \Fig{Fig_2}(b).  
}
\label{Fig_2}
\end{figure}

This complementarity between work and time 
suggests considering the product $W \cdot \tau_{\text{rel}} $ -- rather than either of the two -- as the figure of merit assessing the energy-time cost of a computation. 
On one hand, driven by the demand for speed, effort has been put in pursuing fast-switching storage devices. This has successfully produced state-of-the-art systems with picosecond timescales, though operating far $(\gtrsim 10^6)$ above the reversible limit \cite{Gerrits2002, Zutic2004, Ostler2012, Yang2016}.    
On the other hand, reducing $W$ down to the Landauer limit, at the expense of slow operation,
has been beautifully demonstrated using small particles in traps \cite{Jun2014, Berut2012} or single-domain nanomagnets \cite{Hong2016} as envisioned by Landauer and Bennett \cite{Landauer1961, Bennett1982}.  

In our experiment, a crystal of Fe$_8$ molecular magnets (MM) is used as a spin memory to perform a quantum-enhanced erasure-storage protocol, shown in \Fig{Fig_1}(a). We encode the bit in the "up" and "down" spin states of the MM (\Fig{Fig_1}(b)) and measure the magnetic susceptibility along the erasure-storage cycle. We find that the net work applied to the spin system during this cycle reaches the Landauer limit. This minimal energy cost is retained up to high operation speeds thanks to the possibility of enhancing quantum tunneling of magnetization (QTM) via suitably oriented external magnetic fields.

Each individual  Fe$_8$ molecule represents a magnetic bit and is composed of eight spin-$\frac{5}{2}$ $\text{Fe}^{3+}$-ions coupled to each other by competing antiferromagnetic interactions to form a collective $S=10$ $(20 \mu_\text{B})$ giant-spin. 
By bottom-up chemical synthesis, arrays of these MMs, with perfectly aligned magnetic axes, are packed into a single crystal. 
Due to the relatively large intermolecular spacing,  
the exchange interactions between the molecules are negligible \cite{Burzuri2011}.
The giant-spin $S = 10$ multiplet of the single MM  
is described by the following Hamiltonian \cite{Gatteschi2006}:
\begin{equation}
\mathcal{H} = - DS_z^2 + E(S_x^2 - S_y^2) - g\mu_B \vec S \cdot \vec B.  
\label{eq_1}
\end{equation}
The ligand field, parametrised by the anisotropy constants $D=0.294$ K and $E=0.04$ K, defines $x$, $y$ and $z$ as the hard, medium and easy magnetic axes, respectively, and creates an effective energy barrier with activation energy $U = 26.75$ K \cite{Sangregorio1997} separating the two $S_{z}=\pm 10$ ground eigenstates (Fig. (1b)). The relaxation over this barrier follows approximately Arrhenius' law \taurel$~= \tau_0\exp{(U/k_\text{B}T)}$, where $\tau_0 = 1.43\cdot 10^{-8}$ s is the attempt time \cite{Sangregorio1997}.
{The action of the magnetic field reflects into the Zeeman term of \Eq{eq_1} and is depicted in \Fig{Fig_2}(a), where classical potential and quantum energy levels are represented. A magnetic field parallel to the easy axis, $H_z$, favours either of the two eigenstates $S_z = \pm 10$, i.e., increases the "up" or "down" polarization. Instead, magnetic fields along the medium axis, $H_y$, give rise to off-diagonal terms that mix "up" and "down" states \cite{Burzuri2013}.}  
This allows the spins to tunnel through the potential barrier via progressively lower levels, thus leading to a lower effective $U$ and a consequent decrease in \taurel~\cite{Gatteschi2006}.  

After aligning to the principal magnetic axes of crystal (see \SuppInfo B), we apply the sequence of magnetic fields depicted in \Fig{Fig_2}(b) -- comparable to that proposed in \Ref{Lambson2011} for classical magnets -- in order to carry out the erasure-storage operation on our MM.
In step 1, the magnetic field along the medium axis of the MM ($H_y$) is ramped up to 2 T and the spin states are mixed so that the bit is erased. In step 2, $H_z$ is ramped up to 210 mT in the constant $H_y$ field to initialise the spin in the "up" state. In steps 3 and 4, both the magnetic fields are returned to zero in the same order, closing the cycle and completing the storage. 

Throughout the cycle, the complex ac-susceptibility $\chi_z = \chi^\prime_z + i\chi''_z$ along the easy axis is measured with an inductive susceptometer (see methods for details). {$\chi_z$ is proportional to the derivative of the magnetization $\partial M/\partial H_{\text{ac}}$ and is function of the temperature $T$, frequency $\omega$ of the ac-field $H_{\text{ac}}$ and magnetic field vector $\vec H$. Measuring this quantity allows to track the dynamics of the spin system and its relaxation properties. 
In addition, from $\chi^\prime_z$ the magnetization and work can be derived by integrating once and twice with respect to magnetic field.} The work $W$ obtained in this manner quantifies the heat dissipated during the erasure and measures how reversible the storage operation is \cite{Lambson2011, Hong2016}. 

The experiment is conducted at $T = 1$ K and $\omega/2\pi = 333$ Hz. This temperature is low enough to store the spins for minutes with no field applied, and high enough to have them relaxing within hundreds of nanoseconds when in a transverse field. {Furthermore, at this temperature the ferromagnetic ordering can be neglected  since the dipolar interaction strength is about 0.6 K \cite{Burzuri2011}.}   
The results for the real component of the susceptibility, $\chi'_z$, are shown in \Fig{Fig_2}(c). 
The susceptibility, initially zero, steeply increases at $H_y \approx 0.6$ T, reaches a peak at $H_y \approx 1$ T and slowly decreases up to $H_y  = 2$ T. 
As a small longitudinal field is applied (step 2), $\chi'_z$ sharply drops and reaches zero at $H_z \approx 0.19$ T. Upon retracting the fields in steps 3 and 4, $\chi'_z$ remains substantially zero. 

At the beginning of step 1, the spins are all blocked in either of the potential wells and the wave-function is confined to the "up" or "down" spin eigenstate. 
Upon ramping up $H_y$, 
the admixture of the pure $S_z$ eigenstates is enhanced and the wave-functions delocalised over the two potential wells \cite{Luis1998}: {$\chi^{\prime}_{z}$ increases as the spin is free to tunnel between the two energetically equivalent spin states with the characteristic time \taurel. In the presence of an ac-drive, the spins follow the oscillations of $H_\text{ac}$ provided $\omega \lesssim 1/$\taurel.} 
At this point, the bit is erased. In step 2, a small $H_z$ bias magnetic field is applied. {The susceptibility decreases for increasing spin polarisation reaching zero upon saturation of the magnetization. At this point, the bits are initialised in the "up" configuration.
In step 3, the admixing field is ramped down and thus QTM gradually turned off. This causes the spins to remain \textit{frozen} (i.e., out of equilibrium) in the chosen configuration upon retraction of the bias polarizing field (step 4).}  

\begin{figure}
\includegraphics[width=.99\columnwidth]{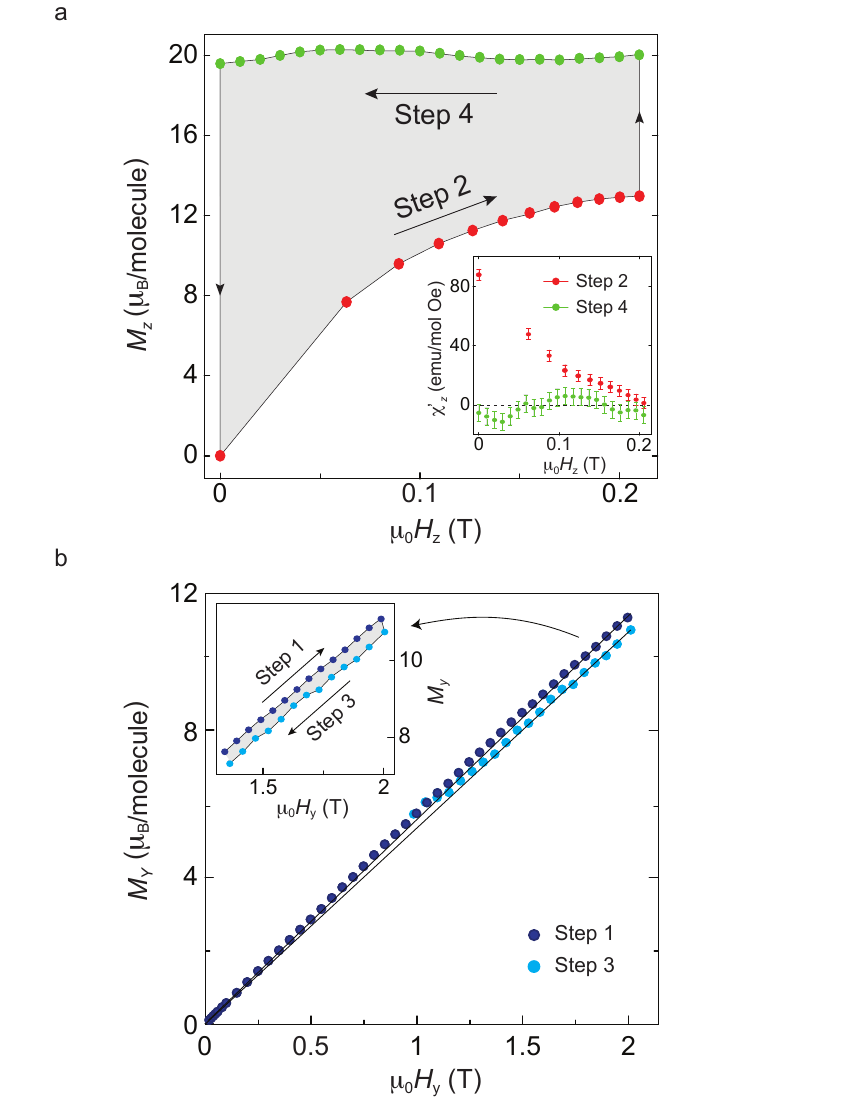}
  \caption{
  \textbf{Total bit storage work.}
	(a) Longitudinal magnetization per molecule $M_z$ extracted by integrating $\chi'_{z}$ of steps 2 and 4 (zoom-in from \Fig{Fig_2}(b) in the inset) with respect to $H_z$. The shaded area corresponds to the work done on the system by $H_z$.	
	(b) SQUID magnetization per molecule $M_y$ along the medium axis of the MM as a function of transverse field $H_y$ at $T = 2$ K. The slope of step 1 is higher than that of step 3 due to the small applied bias $H_z$ field (fits are guidelines to the eye). The area enclosed by these two curves (shaded area in the inset) corresponds to the work done by $H_y$. This work, summed to the one in (a), yields the energy needed for storing a bit of information.}
\label{Fig_3}
\end{figure}

By integrating the measured $\chi'_z$ with respect to $H_z$ (inset of \Fig{Fig_3}(a)), we calculate the easy axis magnetization $M_z$ for steps 2 and 4 (see \SuppInfo A). 
The result is plotted in \Fig{Fig_3}(a). During step 2, $M_z$ increases for increasing $H_y$ before flattening out at about $13\:\mu_B$. Upon retracting $H_y$, $M_z$ increases up to about its maximum value of $20\:\mu_B$ and remains approximately constant as $H_z$ is also ramped to zero. 
The area enclosed by the magnetization loop amounts to the work made by the external magnetic field onto the molecular magnet. 
This yields the value 
$W_{2,4} \equiv W_4 - W_2 = (1.74761 \pm 0.28107)\cdot10^{-16}$ erg/molecule, where the uncertainty corresponds to a 1$\sigma$ confidence interval (see \SuppInfo C for the determination of molecules' number and \SuppInfo D for the uncertainties). To this quantity, the work $W_{1,3}$ done by the $H_y$ in steps 1 and 3 needs to be added. {This work cannot be extracted from the measured medium-axis susceptibility since, due to the strong magnetic anisotropy, $\chi_{y} (\ll \chi_{z})$ is below our detection limit. However, because $\chi_{y}$ is approximately independent of $T$ and $\omega$ (see \SuppInfo B), the magnetization can be invariably measured with a SQUID at $T = 2$ K.}
The result is shown in \Figure{Fig_3}(b). The transverse magnetization $M_y$ is recorded as $H_y$ is ramped up to 2 T (step 1) and subsequently ramped down in the small bias longitudinal field of $0.21$ T (step 3). $M_y$ increases during step 1 and decreases, with a slightly lower slope, during step 3. 
The net work is given by the difference between the works $W_{3}$ and $W_{1}$ done by $H_y$ and amounts to $W_{1,3} \equiv W_{3} - W_{1} = (- 5.6481 \pm 1.7712) \cdot10^{-17}$ erg/molecule. The total dissipated energy, $W$, is then the sum $W = W_{1,3} + W_{2,4} = (1.1828 \pm 0.3322) \cdot10^{-16}$ erg/molecule. Within the experimental uncertainty, this is equivalent to the theoretical Landauer limit at the experimental temperature of 1 K, equal to \landauer $ = 0.9570 \cdot10^{-16} $ erg/bit. This proves that the present system behaves effectively like an ideal "single-spin" bit \cite{Hong2016}. 

\begin{figure}
	\centering
	\includegraphics[width=.99\columnwidth]{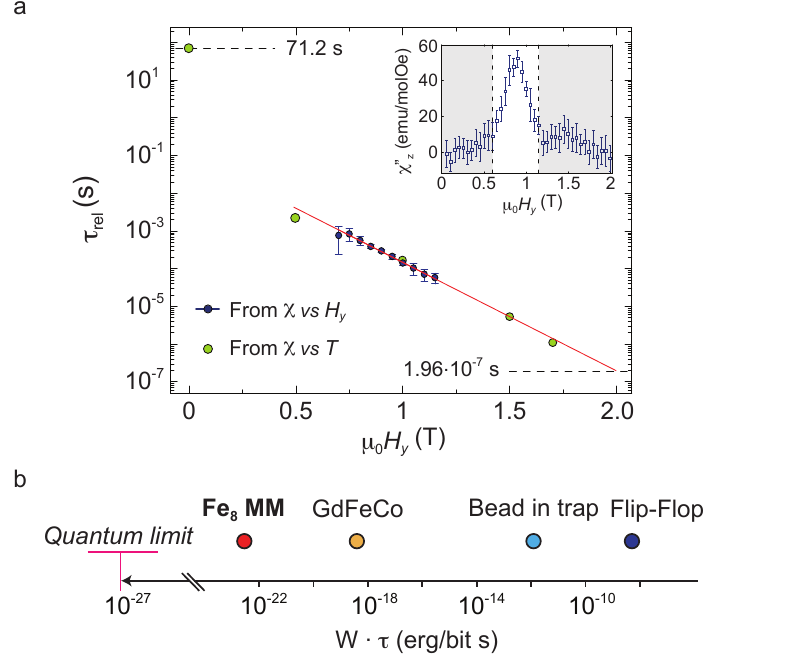}
	\caption{
		\textbf{Relaxation time and energy-time cost.}
		(a) Evolution of the spin relaxation time as a function of $H_y$ during step 1. {The blue data points are obtained from $\chi_z''$ in the interval $0.7 \leq H_y \leq 1.15$ T (inset). The green data points are extracted from temperature sweeps (see \SuppInfo B). The relaxation time reaches 196 ns at $H_y = 2$ T}. This value sets the maximum speed up to which quasi-static operation is retained. (b) Chart comparing the energy-time cost of a storage operation performed with various systems at their respective operating temperature. The Fe$_8$ in this study is the closest to the quantum limit.  
	}
	\label{Fig_4}
\end{figure}

We now discuss the extraction of the magnetic relaxation time. Ac-susceptibility measurements allow for an estimation of the dynamics of the spin relaxation processes. In particular, the ratio 
$\chi_z''/(\omega \chi_z')$ measures the magnetic relaxation time \taurel~or, alternatively said, the time the system takes to reach equilibrium \cite{Gatteschi2006}.   
{In \Fig{Fig_4}(a) we show the evolution of \taurel~as a function of $H_y$ during step 1 in the range $0.7 \leq H_y \leq 1.15$ T for which $\chi_z'' \pm \sigma_{\chi''} \geq 0$ (Inset of \Fig{Fig_4}(a)). We complement these data with \taurel~extracted from $\chi_z'$ measurements in temperature (see \SuppInfo B). The relaxation time exponentially drops from 71.2 s at $H_y = 0$~T to 1.09~$\mu$s at $H_y = 1.7$~T. Extrapolation to $H_y = 2$~T leads to a relaxation time of 196 ns.}
This time is to be interpreted as the longitudinal response time of the phonon bath-and-molecule system upon a change in $H_z$ and fixes the limit up to which quasi-static operation is retained and unwanted (dissipative) hysteresis are avoided. 
The product of the work and relaxation time, $W \cdot \tau_{\text{rel}}$, yields $2.31\cdot10^{-23}$ erg/bit$\cdot$s. This figure quantifies the overall energy-time cost of a computation and its value can be compared to that of other storage devices, operating at room temperature ($T \approx 300$ K). As shown in \Fig{Fig_4}(b) (see \SuppInfo E for the extended chart), the product $W \cdot \tau_{\text{rel}}$ for standard flip-flops -- moderately fast but lossy -- is $\sim 10^{-9}$ erg/bit$\cdot$s; the optical trap system in \Ref{Berut2012}  -- slow but efficient -- attains $\sim 10^{-12}$ erg/bit$\cdot$s. 
Increased performances ($\sim 10^{-19}$ erg/bit$\cdot$s) over these two system is given by the recent GdFeCo laser-driven ferromagnetic element in \Ref{Yang2016} owing to its tens of ps operation time. {The Fe$_8$ MM performs about $10^4$ times better than this system -- reducing to 100 times when accounting for the lower operating temperature.}
Ultimately, the product $W \cdot \tau_{\text{rel}}$ is limited by the Heisenberg uncertainty relation \cite{Margolus1998, Lloyd2000}. According to it, the evolution between two orthogonal, thus classically distinguishable, bit states separated by an energy $\Delta$ would take the minimal "relaxation" time $\tau_{\text{rel}} = \pi \hbar/(2\Delta)$ called quantum speed limit \cite{Aharonov1961, Anandan1990, Deffner2013}. From it, the limit $W \cdot \tau_{\text{rel}} = \pi \hbar/2 = 1.65\cdot10^{-27}$ erg/bit$\cdot$s is obtained as the best trade-off between speed and energy cost \cite{Campbell2017}. Although still far from this limit, the quantum dynamics of systems like the MM in this study proves to be the key to operate both fast and at the Landauer limit and can be used to explore connection between this classical limit and the quantum speed limit.\\


The research reported here was supported by an advanced ERC grant (Mols@Mols). We also acknowledge financial support by the Dutch Organization for Fundamental research (NWO/FOM). 
EB acknowledges funds from the EU FP7 program through the project 618082 ACMOL.
FL acknowledges the Spanish MINECO (grant MAT2015-68204-R), the Gobierno de Arag\'{o}n (grant E98-MOLCHIP) and the European Union (COST 15128 Molecular Spintronics project). R.G. especially thanks Prof. Dr. Luca Gammaitoni for inspiring discussions. 
%
%
%
%


\begin{thebibliography}{10}
\expandafter\ifx\csname url\endcsname\relax
  \def\url#1{\texttt{#1}}\fi
\expandafter\ifx\csname urlprefix\endcsname\relax\def\urlprefix{URL }\fi
\providecommand{\bibinfo}[2]{#2}
\providecommand{\eprint}[2][]{\url{#2}}

\bibitem{Landauer1961}
\bibinfo{author}{Landauer, R.}
\newblock \bibinfo{title}{Irreversibility and heat generation in the computing
  process}.
\newblock \emph{\bibinfo{journal}{IBM Journal of Research and Development}}
  \textbf{\bibinfo{volume}{5}}, \bibinfo{pages}{183--191}
  (\bibinfo{year}{1961}).

\bibitem{Bennett1982}
\bibinfo{author}{Bennett, C.~H.}
\newblock \bibinfo{title}{The thermodynamics of computation - a review}.
\newblock \emph{\bibinfo{journal}{International Journal of Theoretical
  Physics}} \textbf{\bibinfo{volume}{21}}, \bibinfo{pages}{905--940}
  (\bibinfo{year}{1982}).

\bibitem{Bennett1988}
\bibinfo{author}{Bennett, C.~H.}
\newblock \bibinfo{title}{Notes on the history of reversible computation}.
\newblock \emph{\bibinfo{journal}{IBM Journal of Research and Development}}
  \textbf{\bibinfo{volume}{32}}, \bibinfo{pages}{16--23}
  (\bibinfo{year}{1988}).

\bibitem{Leff1990}
\bibinfo{author}{Leff, H.} \& \bibinfo{author}{Rex, A.}
\newblock \bibinfo{title}{Maxwell's demon: Information, entropy, computing}.
\newblock \emph{\bibinfo{journal}{A Hilger and Princeton Univ. Press,
  Europe/USA}}  (\bibinfo{year}{1990}).

\bibitem{Lambson2011}
\bibinfo{author}{Lambson, B.}, \bibinfo{author}{Carlton, D.} \&
  \bibinfo{author}{Bokor, J.}
\newblock \bibinfo{title}{Exploring the thermodynamic limits of computation in
  integrated systems: Magnetic memory, nanomagnetic logic, and the landauer
  limit}.
\newblock \emph{\bibinfo{journal}{Phys. Rev. Lett.}}
  \textbf{\bibinfo{volume}{107}}, \bibinfo{pages}{010604}
  (\bibinfo{year}{2011}).

\bibitem{Gammaitoni2015}
\bibinfo{author}{Gammaitoni, L.}, \bibinfo{author}{Chiuchiu, D.},
  \bibinfo{author}{Madami, M.} \& \bibinfo{author}{Carlotti, G.}
\newblock \bibinfo{title}{Towards zero-power ict}.
\newblock \emph{\bibinfo{journal}{Nanotechnology}}
  \textbf{\bibinfo{volume}{26}}, \bibinfo{pages}{222001}
  (\bibinfo{year}{2015}).

\bibitem{Gerrits2002}
\bibinfo{author}{Gerrits, T.}, \bibinfo{author}{van~den Berg, H. A.~M.},
  \bibinfo{author}{Hohlfeld, J.}, \bibinfo{author}{Bar, L.} \&
  \bibinfo{author}{Rasing, T.}
\newblock \bibinfo{title}{Ultrafast precessional magnetization reversal by
  picosecond magnetic field pulse shaping}.
\newblock \emph{\bibinfo{journal}{Nature}} \textbf{\bibinfo{volume}{418}},
  \bibinfo{pages}{509--512} (\bibinfo{year}{2002}).

\bibitem{Zutic2004}
\bibinfo{author}{\ifmmode \check{Z}\else \v{Z}\fi{}uti\ifmmode~\acute{c}\else
  \'{c}\fi{}, I.}, \bibinfo{author}{Fabian, J.} \& \bibinfo{author}{Das~Sarma,
  S.}
\newblock \bibinfo{title}{Spintronics: Fundamentals and applications}.
\newblock \emph{\bibinfo{journal}{Rev. Mod. Phys.}}
  \textbf{\bibinfo{volume}{76}}, \bibinfo{pages}{323--410}
  (\bibinfo{year}{2004}).

\bibitem{Ostler2012}
\bibinfo{author}{Ostler, T.~A.} \emph{et~al.}
\newblock \bibinfo{title}{Ultrafast heating as a sufficient stimulus for
  magnetization reversal in a ferrimagnet}.
\newblock \emph{\bibinfo{journal}{Nature Communications}}
  \textbf{\bibinfo{volume}{3}}, \bibinfo{pages}{666} (\bibinfo{year}{2012}).

\bibitem{Yang2016}
\bibinfo{author}{Yang, Y.} \emph{et~al.}
\newblock \bibinfo{title}{Ultrafast magnetization reversal by picosecond
  electrical pulses}.
\newblock \emph{\bibinfo{journal}{arXiv preprint arXiv:1609.06392}}
  (\bibinfo{year}{2016}).

\bibitem{Jun2014}
\bibinfo{author}{Jun, Y.}, \bibinfo{author}{Gavrilov, M.} \&
  \bibinfo{author}{Bechhoefer, J.}
\newblock \bibinfo{title}{High-precision test of landauer's principle in a
  feedback trap}.
\newblock \emph{\bibinfo{journal}{Phys. Rev. Lett.}}
  \textbf{\bibinfo{volume}{113}}, \bibinfo{pages}{190601}
  (\bibinfo{year}{2014}).

\bibitem{Berut2012}
\bibinfo{author}{Berut, A.} \emph{et~al.}
\newblock \bibinfo{title}{Experimental verification of landauer's principle
  linking information and thermodynamics}.
\newblock \emph{\bibinfo{journal}{Nature}} \textbf{\bibinfo{volume}{483}},
  \bibinfo{pages}{187--U1500} (\bibinfo{year}{2012}).

\bibitem{Hong2016}
\bibinfo{author}{Hong, J.}, \bibinfo{author}{Lambson, B.},
  \bibinfo{author}{Dhuey, S.} \& \bibinfo{author}{Bokor, J.}
\newblock \bibinfo{title}{Experimental test of landauer'€™s principle in
  single-bit operations on nanomagnetic memory bits}.
\newblock \emph{\bibinfo{journal}{Science advances}}
  \textbf{\bibinfo{volume}{2}}, \bibinfo{pages}{1501492}
  (\bibinfo{year}{2016}).

\bibitem{Burzuri2011}
\bibinfo{author}{Burzur\'{\i}, E.} \emph{et~al.}
\newblock \bibinfo{title}{Magnetic dipolar ordering and quantum phase
  transition in an ${\mathrm{fe}}_{8}$ molecular magnet}.
\newblock \emph{\bibinfo{journal}{Phys. Rev. Lett.}}
  \textbf{\bibinfo{volume}{107}}, \bibinfo{pages}{097203}
  (\bibinfo{year}{2011}).

\bibitem{Gatteschi2006}
\bibinfo{author}{Gatteschi, D.}, \bibinfo{author}{Sessoli, R.} \&
  \bibinfo{author}{Villain, J.}
\newblock \emph{\bibinfo{title}{Molecular nanomagnets}},
  vol.~\bibinfo{volume}{5} (\bibinfo{publisher}{Oxford University Press on
  Demand}, \bibinfo{year}{2006}).

\bibitem{Sangregorio1997}
\bibinfo{author}{Sangregorio, C.}, \bibinfo{author}{Ohm, T.},
  \bibinfo{author}{Paulsen, C.}, \bibinfo{author}{Sessoli, R.} \&
  \bibinfo{author}{Gatteschi, D.}
\newblock \bibinfo{title}{Quantum tunneling of the magnetization in an iron
  cluster nanomagnet}.
\newblock \emph{\bibinfo{journal}{Phys. Rev. Lett.}}
  \textbf{\bibinfo{volume}{78}}, \bibinfo{pages}{4645--4648}
  (\bibinfo{year}{1997}).

\bibitem{Burzuri2013}
\bibinfo{author}{Burzur\'{\i}, E.} \emph{et~al.}
\newblock \bibinfo{title}{Quantum interference oscillations of the
  superparamagnetic blocking in an ${\mathrm{fe}}_{8}$ molecular nanomagnet}.
\newblock \emph{\bibinfo{journal}{Phys. Rev. Lett.}}
  \textbf{\bibinfo{volume}{111}}, \bibinfo{pages}{057201}
  (\bibinfo{year}{2013}).

\bibitem{Luis1998}
\bibinfo{author}{Luis, F.}, \bibinfo{author}{Bartolome, J.} \&
  \bibinfo{author}{Fernandez, J.~F.}
\newblock \bibinfo{title}{Resonant magnetic quantum tunneling through thermally
  activated states}.
\newblock \emph{\bibinfo{journal}{Physical Review B}}
  \textbf{\bibinfo{volume}{57}}, \bibinfo{pages}{505--513}
  (\bibinfo{year}{1998}).

\bibitem{Margolus1998}
\bibinfo{author}{Margolus, N.} \& \bibinfo{author}{Levitin, L.~B.}
\newblock \bibinfo{title}{The maximum speed of dynamical evolution}.
\newblock \emph{\bibinfo{journal}{Physica D}} \textbf{\bibinfo{volume}{120}},
  \bibinfo{pages}{188--195} (\bibinfo{year}{1998}).

\bibitem{Lloyd2000}
\bibinfo{author}{Lloyd, S.}
\newblock \bibinfo{title}{Ultimate physical limits to computation}.
\newblock \emph{\bibinfo{journal}{Nature}} \textbf{\bibinfo{volume}{406}},
  \bibinfo{pages}{1047--1054} (\bibinfo{year}{2000}).

\bibitem{Aharonov1961}
\bibinfo{author}{Aharonov, Y.} \& \bibinfo{author}{Bohm, D.}
\newblock \bibinfo{title}{Time in the quantum theory and the uncertainty
  relation for time and energy}.
\newblock \emph{\bibinfo{journal}{Phys. Rev.}} \textbf{\bibinfo{volume}{122}},
  \bibinfo{pages}{1649--1658} (\bibinfo{year}{1961}).

\bibitem{Anandan1990}
\bibinfo{author}{Anandan, J.} \& \bibinfo{author}{Aharonov, Y.}
\newblock \bibinfo{title}{Geometry of quantum evolution}.
\newblock \emph{\bibinfo{journal}{Phys. Rev. Lett.}}
  \textbf{\bibinfo{volume}{65}}, \bibinfo{pages}{1697--1700}
  (\bibinfo{year}{1990}).

\bibitem{Deffner2013}
\bibinfo{author}{Deffner, S.} \& \bibinfo{author}{Lutz, E.}
\newblock \bibinfo{title}{Energy-time uncertainty relation for driven quantum
  systems}.
\newblock \emph{\bibinfo{journal}{Journal of Physics a-Mathematical and
  Theoretical}} \textbf{\bibinfo{volume}{46}} (\bibinfo{year}{2013}).

\bibitem{Campbell2017}
\bibinfo{author}{Campbell, S.} \& \bibinfo{author}{Deffner, S.}
\newblock \bibinfo{title}{Trade-off between speed and cost in shortcuts to
  adiabaticity}.
\newblock \emph{\bibinfo{journal}{Physical Review Letters}}
  \textbf{\bibinfo{volume}{118}} (\bibinfo{year}{2017}).

\end{thebibliography}

\newpage
\setcounter{figure}{0}    
\onecolumngrid
\appendix
\renewcommand\thefigure{\thesection.\arabic{figure}}  

\section{\textbf{Methods}} 
\noindent \textbf{Susceptibility measurements}
An ac-susceptometer thermally anchored to the mixing chamber of a dilution refrigerator in combination with a 3D vector magnet (9T, 1T, 1T, $0.001^\circ$ accuracy) is used to measure the erasure-storage protocol.
The complex susceptibility \mbox{$\chi (T, \omega) = \chi' (T, \omega) + i\chi'' (T, \omega)$} is measured with a standard lock-in technique with an ac excitation magnetic field of amplitude $H_{ac} = 0.01$ Oe parallel to the common easy axis of the molecules. \\

\noindent \textbf{Magnetization measurements} 
Magnetization is measured with a commercial SQUID magnetometer ($T \geq 1.8 $ K) equipped with a rotating stage and an ac susceptibility option. \\

\noindent \textbf{Calculations of magnetization and work} 
The easy-axis magnetization $M_z$ is obtained from the susceptibility $\chi^{\prime}_z$ by making use of the integral: 
\begin{equation*}
M_z = \int \chi'_z \: dH_z.
\label{eq_2}
\end{equation*}
The works done by $H_z$ (steps 2 and 4) and $H_y$ (steps 1 and 3) are calculated by performing an analogous integration on the resulting $M_z$ and $M_y$, respectively:
\begin{align*}
W_{2,4} = \oint M_z \: dH_z, \:\:\:\: W_{1,3} = \oint M_y \: dH_y. 
\label{eq_3}
\end{align*}
These correspond to the loop shaded areas in \Fig{Fig_3} (a) and (b). 

\section{Characterisation in temperature and transverse field\label{characterisation}}
This section contains details on: (i) the characterisation of the superparamagnetic behaviour of the MM crystal in temperature and frequency; (ii) the extraction of the relaxation time as a function of selected $H_y$ fields; (iii) the determination of the crystal's magnetic easy and medium axes orientation with respect to the laboratory's reference system.  \\

\noindent \textbf{(i) Temperature and frequency}
\begin{figure*}
	\centering
	\includegraphics[width=.68\columnwidth]{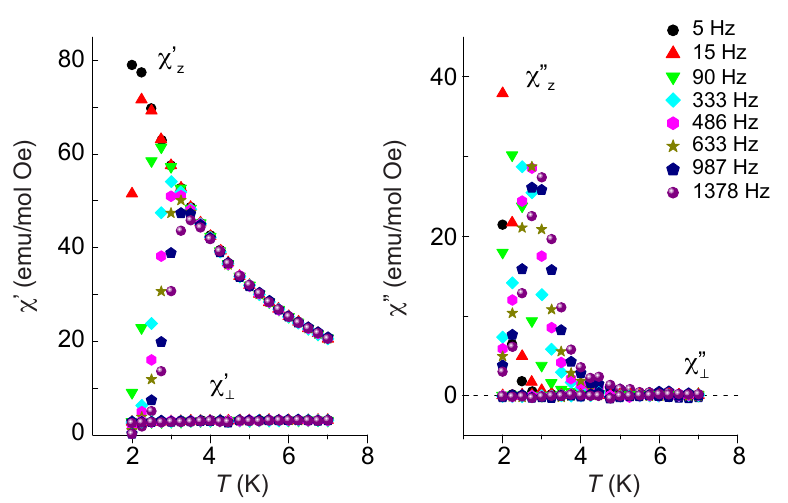}
	\caption{
		\textbf{Temperature and frequency characterization.}
		(a) Real (left) and imaginary (right) part of the susceptibility as a function of $T$ and indicated frequency $\omega$ ranging from 5 Hz to 1378 Hz. The frequency-dependent departure from equilibrium signals the expected superparamagnetic behaviour.   
	}
	\label{Fig_S1}
\end{figure*}
In \FigS{Fig_S1}, the real (in-phase) $\chi' (T, \omega)$ and imaginary (out-of-phase) $\chi'' (T, \omega)$ components of the ac-susceptibility are shown as a function of temperature for the indicated frequencies in zero magnetic field. For each component, the longitudinal $\chi_z$ and transverse $\chi_\perp$ parts are plotted. For a fixed frequency, decreasing the temperature results in the increase of $\chi'_z$ accompanied by a constant $\chi''_z \approx 0$. This is the typical behaviour of a standard paramagnet, where the absence of an out-of-phase response signals equilibrium and a fast relaxation time $\tau_\text{rel} \ll 1/\omega$. 
However, as a frequency-dependent temperature is reached, $\chi'_z$ starts dropping to zero while $\chi''_z$ exhibits a peak. This temperature corresponds to the so-called blocking temperature, $T_b$, and it is characterised by a spin relaxation time $\tau_\text{rel} \simeq 1/\omega$. For $T < T_b$, the spin of the MM is increasingly driven out-of-equilibrium and $\tau_\text{rel}$ further increases. The observed behaviour is a fingerprint of the superparamagnetism expected in a MM, where the potential  barrier prevents fast spin relaxation at sufficiently low temperatures.   
The small temperature- and frequency-independent $\chi'_\perp$ and zero $\chi''_\perp$ further signal the strong spin polarisation along the easy-axis and negligible transverse (hard-plane) spin projection. 

\newpage

\begin{figure}[H]
	\centering
	\includegraphics[width=.51\columnwidth]{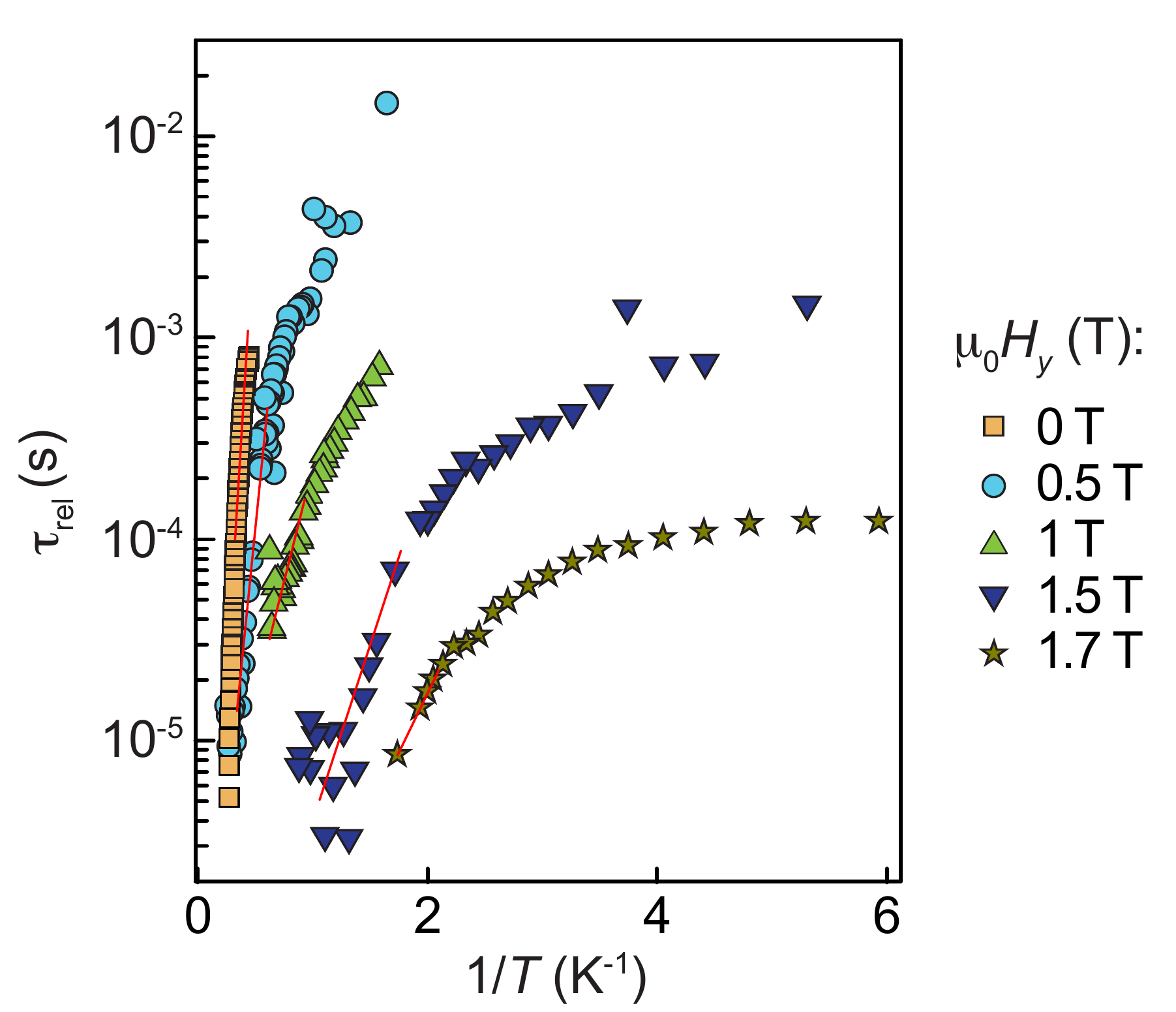}
	\caption{
		\textbf{Relaxation time for different transverse fields.}
		(a) $\tau_\textit{rel}$ obtained from $\chi''_z/\omega\chi'_z$ as a function of inverse temperature. Fitting of the temperature-dependent side allows to obtain the desired $\tau_\textit{rel}(H_y)$ at $T= 1$ K by extrapolation.  
	}
	\label{Fig_S2}
\end{figure}

\noindent \textbf{(ii) Transverse magnetic field}
To extract the relaxation time data-points shown in Fig. 4 (and labelled "$\chi$ {\em vs} $T$ data"), we use temperature-dependent complex susceptibility measurements for the different $H_y$ fields at the frequency $\omega = 333$ Hz. Plotting the ratio $\chi''_z/\omega\chi'_z~=\tau_\textit{rel}$ as a function of the inverse temperature results in \FigS{Fig_S2}.
At high temperatures, the relaxation time behaves according to Arrhenius' law $\log\tau_\textit{rel}~= U_\textit{eff}/T + \tau_0$ with the effective barrier $U_\textit{eff}(H_y)$ obtained by fitting the temperature-dependent part of the curves. Extrapolation of the fit to $T = 1$ K, yields $\tau_\textit{rel}(H_y; T = 1 \:\text{K})$ for the selected fields. 
\noindent \textbf{(iii) Alignment to principal axes}
In this subsection we describe the procedure used for finding the accurate orientation of the principal axes of the MM with respect to the $X$, $Y$ and $Z$ axes of the vector magnet. Provided this orientation is approximately known, the crystal is placed in the susceptometer with its easy, medium and hard axes about the $Y$, $Z$ and $X$-axis of the magnet, respectively. Measurements are executed at $T = 3$ K and $\omega = 1333$ Hz. Under these conditions, the susceptibility is close to equilibrium (see \FigS{Fig_S1}) and thus strongly dependent on the magnetic field orientation. The first operation consists of rotating the magnetic field $\mu_0|H| = 0.1$ T on the $XZ$-plane by fixing $\phi = 0$ and sweeping $\theta$. 
\begin{figure*}
	\centering
	\includegraphics[width=.99\columnwidth]{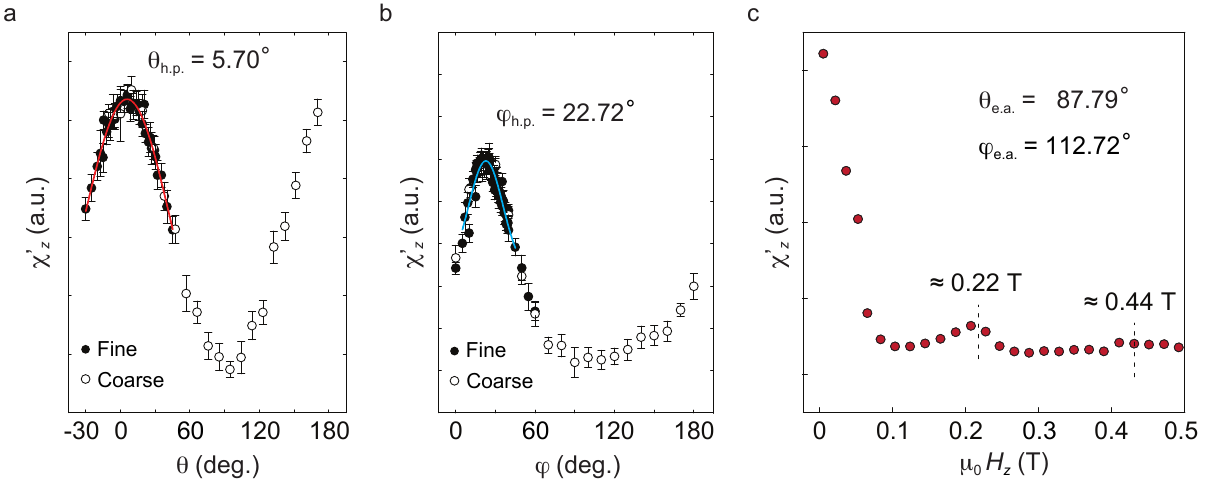}
	\caption{
		\textbf{Determination of the easy axis.}
		(a) $\chi'_z$ as a function of angle $\theta$ for fixed $\phi = 0^\circ$ ($XZ$-plane). (b) Same as (a) for fixed $\theta = 90^\circ$ ($XY$-plane). Maxima signal the two crossings with the hard plane from which the orientation of the easy axis is obtained. (c) $\chi'_z$ as a function of magnetic field intensity along the easy axis (labelled by the subscript $z$). The two peaks signal the expected magnetic level crossings.          
	}
	\label{Fig_S3}
\end{figure*}
\begin{figure*}
	\centering
	\includegraphics[width=.4\columnwidth]{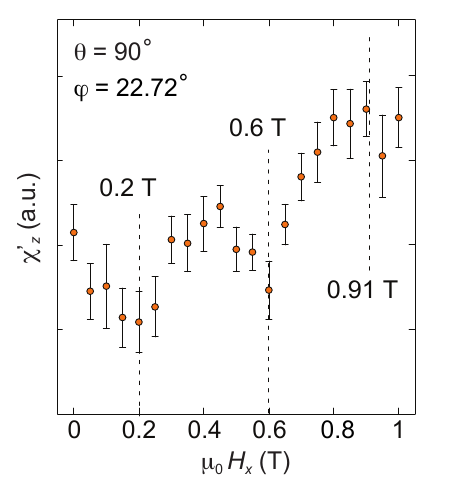}
	\caption{
		\textbf{Quantum interference pattern along the hard axis.}
		$\chi'_z$ as a function of magnetic field intensity along the hard axis (labelled by the subscript $x$). Oscillations signal the expected quantum interference pattern with minima at the indicated fields. 
	}
	\label{Fig_S32}
\end{figure*}
As \FigS{Fig_S3}(a) shows, $\chi'_z$ exhibits a maximum (minimum), signalling a condition relatively closer to (further from) equilibrium. In correspondence of the maximum, for $\theta = 5.70^\circ$, the magnetic field crosses the hard plane, while it is closest to the easy axis in correspondence of the minimum. An analogous operation is conducted on the $XY$-plane (\FigS{Fig_S3}(b)), where the crossing with the hard plane occurs for $\phi = 22.72^\circ$. The cross product between the two hard plane vectors yields an easy axis with angular coordinates $\theta = 87.79^\circ$ and $\phi = 112.72^\circ$. This axis is hereafter labelled by the subscript $z$. A confirmation of the accurate orientation of this axis is shown in \FigS{Fig_S3}(c). Sweeping the magnetic field along it gives rise to a peak at 0.22 T (0.44 T), in correspondence of the resonance between the spin eigenstates $S_z = 10$ and $S_z = -9$ ($S_z = 10$ and $S_z = -8$) expected at $B_n = \frac{D}{g\mu_\text{B}}n = 0.219\:\text{T} \times n$, for $n=1\: (2)$. 

At this point, the magnetic field is swept on the hard plane for $\theta = 90^\circ$ and $\phi = 22.72^\circ$ (\FigS{Fig_S32}). The observed oscillatory behaviour in $\chi'_z$, with minima at the indicated fields, is in accordance with the characteristic quantum interference pattern (see Ref. [17] of the main text) occurring in the proximity to the hard axis -- labelled by $x$ hereafter. A $90^\circ$-shift from this axis along the hard plane fixes the medium axis -- labelled by $y$ -- and concludes the orientation procedure.

\section{Determination of the number of molecules}\label{number}
\begin{figure}[H]
	\centering
	\includegraphics[width=.7\columnwidth]{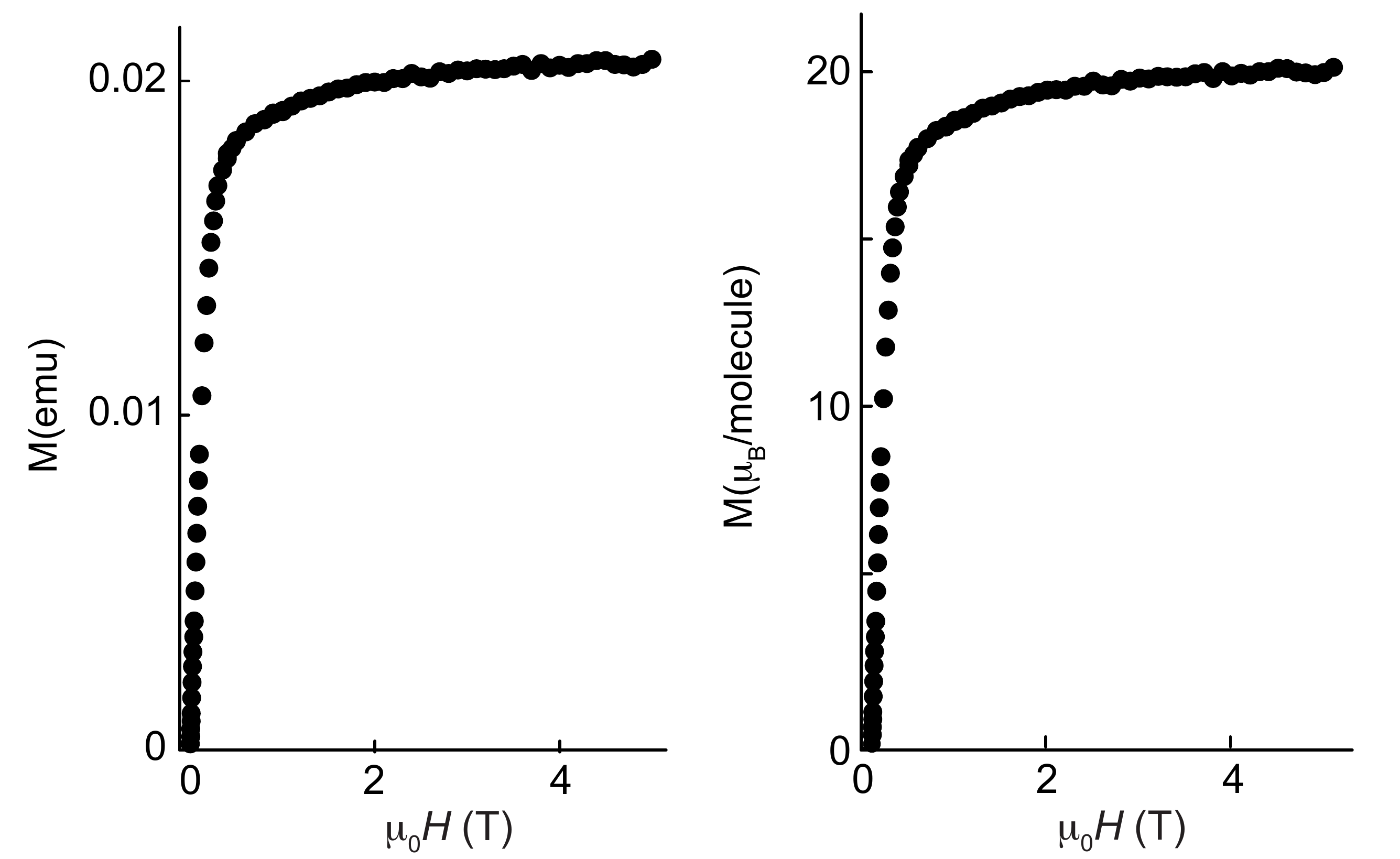}
	\caption{
		\textbf{Saturation of the magnetization.}
		(a) Raw magnetization $M$ (emu) measured in the SQUID as a function of magnetic field. 
		(b) Scaled magnetization $M (\mu_\text{B}/$molecule) obtained normalizing the raw magnetization to the single-molecule value of 20 $\mu_\text{B}$. The ratio between the two quantities yields the indicated estimate for the number of molecules in the crystal.    
	}
	\label{Fig_S4}
\end{figure}

We have determined the number of molecules (bits) in the crystal with two independent methods. The first and most straightforward is that of dividing the weight of the crystal, $m = 0.411$ mg, by the molecular weight, $P_\text{m} = 2262.45$ g/mol and multiply by the Avogadro constant. This yields a number of molecules $N = 1.094 \cdot 10^{17} $. 
The second method takes advantage of the fact that each molecule has a definite spin $S = 10$ $(20 \mu_\text{B})$. By measuring the saturation magnetization, $M_s$, of the crystal in the SQUID setup and dividing by the spin of the single molecule yields: 
\begin{align*}
N = \frac{M_s (\text{emu})}{20 \mu_\text{B}}\:\:5.1883 \cdot 10^{20}\:\mu_\text{B}/\text{emu}
\end{align*}   
Provided $M_s = (2.029 \pm 0.006)\cdot 10^{-2}$ emu (\FigS{Fig_S4}), $N = (1.09392 \pm 0.00326)\cdot 10^{17}$. 
\section{Uncertainty estimation\label{errors}}
The values of the susceptibility $\chi^{\prime}_{z}$ given in Fig.2 at each magnetic field result from averaging over $n = 15$ samples. The uncertainty on the mean, $\sigma_\chi$, is calculated as its standard deviation assuming a normal distribution. The magnetization per molecule $M_z$ is a function of $\chi^{\prime}_{z}$ and the number of molecules $N$ and is therefore affected by an uncertainty $\sigma_M (H_z)$ given by the propagated uncertainties: 
\begin{align*}
\sigma_M (H_z) = \sqrt{\left(\frac{\partial M_z}{\partial \chi^{\prime}_{z}}\right)^2 \sigma_\chi^2 + \left(\frac{\partial M_z}{\partial N} \right)^2 \sigma_N^2}. 
\end{align*}   
Since the work $W_{2,4}$  is calculated as:
\begin{align*}
W_{2,4} = \oint M_z(H_z) \: dH_z,
\end{align*}
its upper ($+$) and lower ($-$) confidence bounds are given by: 
\begin{align*}
W_{2,4}^\pm = \oint (M_z (H_z) \pm \sigma_M (H_z)) \: dH_z, 
\end{align*}
So that its associated standard deviation $\sigma_{2,4}$ is: 
\begin{align*}
\sigma_{2,4} = |W_{2,4} - W_{2,4}^\pm|.
\end{align*}

The uncertainty, $\sigma_{1,3}$, on the work $W_{1,3}$ -- obtained by integrating $M_y$ -- is calculated using an analogous procedure.

The total error affecting the work $W = W_{1,3} + W_{2,4}$ is then $\sigma = \sqrt{\sigma_{1,3}^2 + \sigma_{2,4}^2}$. 

\section{Details on the energy-time cost of a computation\label{product}}
\begin{figure}[H]
	\centering
	\includegraphics[width=.55\columnwidth]{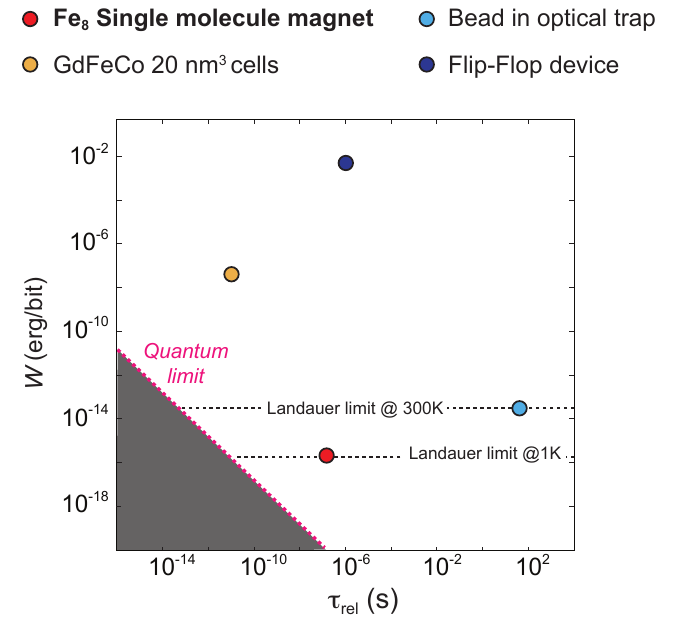}
	\caption{
		\textbf{Relaxation time and efficiency comparison.}
		(a) Chart comparing the energy-time cost of a storage operation performed with various systems. The Fe$_8$ is the closest to the quantum limit.  
	}
	\label{Fig_S5}
\end{figure}

Here we report a more complete version of the chart in figure 4 of the main text where the quantities $W$ and $\tau$ determining the product $W \cdot \tau$ are plotted on the Cartesian plane (\FigS{Fig_S5}). 
Along the diagonal line, at the top-right corner of the plane sit slow and lossy devices whereas the fast and efficient ones are on the bottom-left. The devices belonging to this corner of the plane are bounded from below by the quantum limit $W \cdot \tau_{\text{rel}} = \pi \hbar/2 = 1.65\cdot10^{-27}$ erg/bit$\cdot$s. \\

\end{document}